\documentclass[twocolumn]{aa}
\usepackage{graphicx}
\begin{document}
\title{Cosmological effects on the observed flux and
fluence distributions of gamma-ray bursts:\\
Are the most distant bursts in general the faintest ones?}
\titlerunning{Cosmological effects on the observed...}

\author{A. M\'esz\'aros \inst{1}
\and
J. \v{R}\'{\i}pa \inst{1}
\and
F. Ryde \inst{2}}

\offprints{A. M\'esz\'aros}

\institute{Charles University, Faculty of Mathematics and Physics, Astronomical Institute,\\
V Hole\v{s}ovi\v{c}k\'ach 2, CZ 180 00 Prague 8, Czech Republic\\
\email{meszaros@cesnet.cz}\\
\email{ripa@sirrah.troja.mff.cuni.cz}
\and
Department of Physics, Royal Institute of Technology, AlbaNova University Center,\\
SE-106 91 Stockholm, Sweden\\
\email{felix@particle.kth.se}
}

\date{Received May 3, 2010; accepted January 19, 2011}

\abstract {Several claims have been put forward that an essential
fraction of long-duration BATSE gamma-ray bursts should lie at
redshifts larger than 5. This point-of-view follows from the
natural assumption that fainter objects should, on average, lie at
larger redshifts. However, redshifts larger than 5 are rare for
bursts observed by Swift, seemingly contradicting the BATSE
estimates.}{The purpose of this article is to clarify this
contradiction.}{We derive the cosmological relationships between
the observed and emitted quantities, and we arrive at a prediction that can be
tested on the ensembles of bursts with determined redshifts. This
analysis is independent on the assumed cosmology,
on the observational biases, as well as on any
gamma-ray burst model. Four different samples are studied: 8
BATSE bursts with redshifts, 13 bursts with derived
pseudo-redshifts, 134 Swift bursts with redshifts, and 6 Fermi
bursts with redshifts.}{The controversy can be explained by the fact that
apparently fainter bursts need not, in general, lie at large redshifts.
Such a behaviour is possible,
when the luminosities (or emitted energies) in a sample of bursts increase
more than the dimming of the observed values with redshift. In such a case
$dP(z)/dz > 0$ can hold, where $P(z)$ is either the peak-flux or the fluence.
All four different samples of the long bursts suggest
that this is really the case. This also means that the hundreds of faint,
long-duration BATSE bursts need not lie at high redshifts, and
that the observed redshift distribution of long Swift bursts might
actually represent the actual distribution.}{}


\maketitle

\section{Introduction}

Until the last years, the redshift distribution of gamma-ray bursts
(GRBs) has only been poorly known. For example, the Burst and
Transient Source Experiment (BATSE) instrument on Compton Gamma
Ray Observatory detected around 2700 
GRBs\footnote{http://www.batse.msfc.nasa.gov/batse/grb/catalog/}, but only
a few of these have directly measured redshifts from the afterglow
(AG) observations (\cite{scha03,pi04}). During the last couple of
years the situation has improved, mainly due to the observations
made by the Swift 
satellite\footnote{http://swift.gsfc.nasa.gov/docs/swift/swiftsc.html}.
 There are already
dozens of directly determined redshifts (\cite{me06}).
Nevertheless, this sample is only a small fraction of the, in
total, thousands of detected bursts.

Beside the direct determination of redshifts from the AGs, there
are several indirect methods, which utilize the gamma-ray data
alone. In essence, there are two different methods which provide
such determinations. The first one makes only statistical
estimations; the fraction of bursts in a given redshift interval
is deduced. The second one determines an actual value of the
redshift for a given GRB ("pseudo-redshift").

Already at the early 1990's, that is, far before the observation
of any AG, and when even the cosmological origin was in doubt,
there were estimations made in the sense of the first method
(see, e.g., Paczy\'nski (1992) and the references therein). In
M\'esz\'aros \& M\'esz\'aros (1996) a statistical study indicated
that a fraction of GRBs should be at very large redshifts (up to
$z \simeq 20$). In addition, there was no evidence for the
termination of the increase of the numbers of GRBs for $z>5$ (see
also M\'esz\'aros \& M\'esz\'aros (1995); Horv\'ath et al. (1996),
and Reichart \& M\'esz\'aros (1997)). In other words, an essential
fraction (a few or a few tens of~\%) of GRBs should be in the
redshift interval $5 < z < 20$. Again using this type of
estimation, Lin et al. (2004) claims that even the {\it majority}
of bursts should be at these high redshifts.

The estimations of the pseudo-redshifts in the sense of the second
method are more recent. Ramirez-Ruiz \& Fenimore (2000), Reichart
et al. (2001), Schaefer et al. (2001) and Lloyd-Ronning et al.
(2002) developed a method allowing to obtain from the so-called
variability the intrinsic luminosity of a GRB, and then from the
measured flux its redshift. The redshifts of hundred of bursts
were obtained by this method. Nevertheless, the obtained
pseudo-redshifts are in doubt, because there is an $(1+z)$ factor
error in the cosmological formulas (\cite{nor04},
 Band et al. 2004). Other authors
also query these redshifts (e.g., \cite{gui05}). Norris et al.
(2000) found another relation between the spectral lag and the
luminosity. This method seems to be a better one
(\cite{scha01,nor02,ryde05}), and led to the estimation of $\sim
1200$ burst redshifts. Remarkably, again, an essential fraction of
long GRBs should have $z >5$, and the redshift distribution should
peak at $z \sim 10$. A continuation of this method (\cite{band04},
Norris 2004), which corrected the $(1+z)$ error in Norris
(2002), gave smaller redshifts on average, but the percentage of long
GRBs at $z>5$ remains open. Bagoly et al. (2003) developed a different
method allowing to obtain the redshifts from the gamma spectra for
343 bright GRBs. Due to the two-fold character of the estimation,
the fraction of GRBs at $z >5$ remains further open. No doubt has
yet emerged concerning this method. Atteia (2003) also proposed a
method in a similar sense for bright bursts. Other methods of such
estimations also exist (\cite{am02,ghi05}). These pseudo-redshift
estimations (for a survey see, e.g. Sect.2.6 of M\'esz\'aros (2006))
give the results that even bright GRBs should be at redshifts $z
\sim (1-5)$. For faint bursts in the BATSE data set (i.e. for GRBs
with peak-fluxes smaller than $\simeq 1$\,photon/(cm$^2$s)) one
hardly can have good pseudo-redshift estimations, but on average
they are expected to be at larger redshifts, using the natural
assumption that these bursts are observationally fainter due to their
larger distances. Hence, it is remarkable that all these
pseudo-redshift estimations also supports the expectation,
similarly to the results of the first method, that an essential
fraction of GRBs is at very high redshifts.

Contrary to these estimations for the BATSE data set, only five long
bursts with $z > 5$ bursts have yet been detected from direct redshift
measurements from AGs
using more recent satellites. In addition, the majority of
measured $z$-s are around $z = 2-3$, and there is a clear
decreasing tendency towards
the larger redshifts\footnote{http://www.mpe.mpg.de/$\sim$jcg/grbgen.html}. 
In other words, the redshifts
of GRBs detected by the Swift satellite do not support the BATSE
redshift estimations; the redshifts of GRBs detected by the
non-Swift satellites are on average even at smaller redshifts
(\cite{bag06}).

This can be interpreted in two essentially different ways. The
first possibility is that a large fraction (a few tens of~\% or
even the majority) of GRBs are at very high redshifts (at $z
> 5$ or so). In such case these bursts should mainly be seen
only in the gamma-ray band due to some observational selection
(\cite{lare00}). The second possibility is that the AG data
reflect the true distribution of bursts at high redshifts, and
bursts at $z > 5$ are really very rare. In this case, however,
there has to be something wrong in the estimations of
redshifts from the gamma-ray data alone. Since observational
selections for AG detections of bursts at $z >5$ cannot be
excluded (\cite{lare00}), the first point-of-view could be also
quite realistic.

The purpose of this paper is to point out some statistical
properties of the GRBs, which may support the second possibility.
Section 2 discusses these properties theoretically. In sections 3
and 4 we discuss the impact of such a behaviour on several
observed burst samples. Section 5 summarizes the results of paper.

\section{Theoretical considerations}

\subsection{The general consideration}

Using the standard cosmological theory and some simple statistical
considerations, we will now show that,
under some conditions, apparently fainter bursts might very well be at
smaller redshifts compared to the brighter ones.

As shown by M\'esz\'aros \& M\'esz\'aros (1995),
if there are given two photon-energies $E_1$ and
$E_2$, where $E_1 < E_2$, then the flux $F$ (in units photons/(cm$^2$s))
of the photons with  energies $E_1 \leq E \leq E_2$
detected from a GRB having a redshift $z$ is given by
\begin{equation}
F_\mathrm{ph}(z) = \frac{(1+z)
\tilde{L}_\mathrm{ph}(z)}{4\pi d_\mathrm{l}(z)^2} =  \frac{
\tilde{L}_\mathrm{ph}(z)}{4\pi d_\mathrm{M}(z)^2 (1+z)}, \label{eq:1}
\end{equation}
where $\tilde{L}_\mathrm{ph}(z)$ is the isotropic luminosity of a
GRB (in units photons/s) between the
energy range $E_1(1+z) \leq E \leq E_2(1+z)$, and $d_\mathrm{l}(z)$
is the luminosity distance of the source. The reason of
the notation $\tilde{L}(z)$, instead of
the simple $L(z)$, is that $L(z)$ should mean the
luminosity from $E_1 \leq E \leq E_2$ (\cite{me06b}).
One has $d_\mathrm{l}(z) = (1+z) d_\mathrm{M}(z)$,
where $d_\mathrm{M}(z)$ is the proper motion distance of the GRB, given
by the standard cosmological formula (\cite{ca92}),
and depends on the cosmological parameters $H_0$
(Hubble-constant), $\Omega_\mathrm{M}$ (the ratio of
the matter density and the critical density), and $\Omega_\mathrm{\Lambda}
= 3 \lambda c^2/(3 H_0^2)$ ($\lambda$ is the cosmological constant,
$c$ is the velocity of light). In energy units
one may write $F_\mathrm{en} (z) = \overline{E} F_\mathrm{ph}(z)$ and
$\tilde{L}_\mathrm{en}(z) =
(1+z) \overline{E} \tilde{L}_\mathrm{ph}(z)$, where $\overline{E}$ ($E_1 <
\overline{E} <E_2 $) is a typical photon energy ensuring that the
flux $F_\mathrm{en}(z)$ has the dimension erg/(cm$^2$s).
$\tilde{L}_\mathrm{en}(z)$ in
erg/s unit gives the luminosity in the interval $E_1(1+z) \leq E \leq E_2(1+z)$.
Except for an $(1+z)$ factor the situation is the same, when
considering the fluence. Hence, in the general case, we have
\begin{equation}
P(z) = \frac{(1+z)^N \tilde{L}(z)}{4\pi d_\mathrm{l}(z)^2}, \label{eq:2}
\end{equation}
where $P(z)$ is either the fluence or the flux, and $\tilde{L}(z)$ is either
the emitted isotropic total number of photons, or the isotropic
total emitted energy or the luminosity. The following values of
$N$ can be used: $N=0$ if the flux is taken in energy units erg/(cm$^2$s)
and $\tilde{L}$ is the luminosity with dimension erg/s; $N=1$ if
either the flux and luminosity are in photon units, or the fluence
in energy units and the total energy are taken; $N=2$ if
the total number of photons are considered. All this means that
for a given GRB - for which redshift, flux and fluence are
measured - Eq.(\ref{eq:2}) defines immediately $\tilde{L}(z)$,
which is then either the isotropic total emitted energy or the luminosity
in the interval $E_1(1+z) \leq E \leq E_2(1+z)$.
Hence, $\tilde{L}(z)$ is from the range $E_1(1+z) \leq E \leq E_2(1+z)$
and not from $E_1 \leq E \leq E_2$.

Let us have a measured value of $P(z)$. Fixing this $P(z)$  Eq.(\ref{eq:2})
defines a functional relationship between the redshift $z$ and
$\tilde{L}(z)$. For its transformation into
the real intrinsic luminosities $L(z)$ the beaming must be taken into account as
well (\cite{lee00,ryde05,bv09a,pet09,bv09b,bu10}). Additionally,
we need to study the dependence of the obtained  $\tilde{L}(z)$ on $z$, and to determine
the real luminosities $L(z)$ by the K-corrections (\cite{me06}).
It is not the aim of this paper to solve the transformation of
$\tilde{L}(z)$ into $L(z)$. The purpose of this paper is to study the
functional relationship among $P(z)$, $z$ and $\tilde{L}(z)$.

Using the proper motion distance $d_\mathrm{M}(z)$
Eq.(\ref{eq:2}) can be reordered as
\begin{equation}
4\pi d_\mathrm{M}(z)^2 (1+z)^{2-N} P (z) = \tilde{L}(z). \label{eq:3}
\end{equation}
The proper motion distance  $d_\mathrm{M}(z)$ is bounded as $ z
\rightarrow \infty$
(Weinberg 1972, Chapter. 14.4.). This finiteness of the proper
motion distance is
fulfilled for any $H_0,\; \Omega_\mathrm{M}$ and $\Omega_\mathrm{\Lambda}$.
Hence,  $\tilde{L}(z)$ is a monotonically increasing function of the redshift
along with $(1+z)^{2-N}$ for the fixed $P(z) = P_0$ and for the given value of $N \leq 1$.
It means $z_1 < z_2$ implies $\tilde{L}(z_1) < \tilde{L}(z_2)$.
Expressing this result in other words: the more distant and
brighter sources may give the same observed value of $P_0$.
Now, if a source at $z_2$ has a
$\tilde{L}’ > \tilde{L}(z_2)$, its observed value $P'_{obs}$ will have
$P'_{obs} > P_0$, i.e. it becomes apparently brighter despite its
greater redshift than that of the source at $z_1$. The probability of the
occurrence this kind of {\it inversion} depends on $f(\tilde{L}|z)$, on
the conditional probability
density of $\tilde{L}$ assuming $z$ is given,
and on the spatial density of the objects.

It is obvious from  Eq.(\ref{eq:3}) that for the increasing $z$ the
$\tilde{L}(z)$ is also increasing. This effect gives a bias
(\cite{lee00,bv09a,pet09,bv09b,bu10}) towards the higher
$\tilde{L}(z)$ values among GRBs observed above a given detection
threshold. [These questions
are discussed in detail mainly by Petrosian et al. (2009).] There can be
also that $\tilde{L}(z)$ is increasing with $z$ due to the metallicity evolution
(\cite{wp07}). There can be also an intrinsic evolution
of the $L(z)$ itself in the
energy range $[E_1, E_2]$.  Hence, keeping all this in mind, $\tilde{L}(z)$
can well be increasing on $z$, and the inverse behaviour can also occur.

\subsection{A special assumption}

Assume now that we have $\tilde{L}(z) \propto (1+z)^{\mathrm{q}}$, where
$q$ is real number, and this relation holds for any $0 < z < \infty$.
This means that it holds  $\tilde{L}(z) = \tilde{L}_0
(1+z)^{\mathrm{q}}$, where $\tilde{L}(z=0)  = \tilde{L}_0$.
Of course, this special assumption is a loss of generality, because
$\tilde{L}(z)$ can be really a function of $z$, but in general it need not
have this special form. In addition, to calculate $\tilde{L}(z)$ the cosmological
parameters must be specified, which is a further loss of generality.
Nevertheless, if this special assumption is taken,
the inverse behavior may well be demonstrated.

If $(N + q) > 2$, then one has a highly interesting mathematical
behaviour of the function $P(z)$ (Eq.\ref{eq:2}). For $z \ll 0.1 $,
$P(z)$ decreases as $z^{-2}$, that is, larger
redshifts gives smaller fluxes or fluences as expected. However,
after some $z = z_\mathrm{turn}$ this behavior must change, because as
$z$ tends to $\infty$, the function $(1+z)^{\mathrm{N}} \tilde{L}
/d_\mathrm{l}^2$ tends to infinity following $\propto (1+z)^{\mathrm{N + q -
2}}$. In other words, for $z > z_\mathrm{turn}$ {\it as redshift
increases, the measured
$P(z)$ will also increase}. Equivalently stated, "fainter bursts are
closer to us". The possibility of this "inverse" behaviour is
quite remarkable.
It is important to note that the existence of a $z_\mathrm{turn}$ is
exclusively determined by the value $q$, and the necessary and
sufficient condition for it is given by the inequality $(N + q) > 2$.
For the existence of a $z_\mathrm{turn}$ the
values $H_0, \Omega_\mathrm{M}$ and $\Omega_\mathrm{\Lambda}$ are
unimportant. The value of $z_\mathrm{turn}$ can vary depending on the
choice of the
$\Omega$ parameters, but, however, its existence or non-existence is
unaffected.

Moreover, the value of $z_\mathrm{turn}$ itself is independent on the
Hubble-constant $H_0$. This can be seen as follows. To find
$z_\mathrm{turn}$ one must simply search for the minimum of the function
$Q(z) = (1+z)^{N+q}/d_\mathrm{l}(z)^2$, that is,
when $dQ(z)/dz = 0$. But, trivially, $Q(z)$ and $Q(z)/H_0^2$
have the same minimum.

\begin{figure*}[]
\centering
 \includegraphics[width=0.95\textwidth]{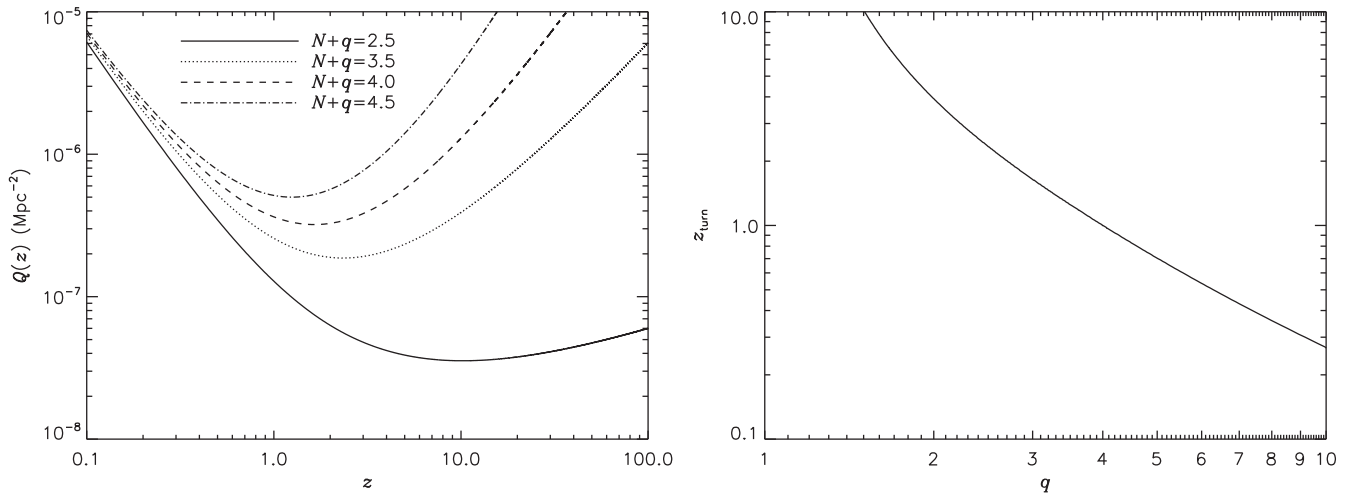}
 \caption{{\it Left panel}: Function $Q(z)$ for $\Omega_\mathrm{M} =
0.27$ and $\Omega_\mathrm{\Lambda} = 0.73$.
{\it Right panel}: Dependence of $z_\mathrm{turn}$ on $q$ for
          $\Omega_\mathrm{M} = 0.27$ and $\Omega_\mathrm{\Lambda} =
0.73$.}
 \label{fig:zturn}
\end{figure*}

The solution of the equation $dQ(z)/dz = 0$ must be found
numerically for the general case of Omega parameters.
The left panel of Fig.\ref{fig:zturn} shows the function $Q(z)$ for
$\Omega_\mathrm{M} = 0.27$ and $\Omega_\mathrm{\Lambda} = 0.73$.
For $\Omega_\mathrm{\Lambda} = 0$
it can be found analytically, because $d_\mathrm{M}(z)$ is then an
analytical function. For $\Omega_\mathrm{M} = 1$ and
$\Omega_\mathrm{\Lambda} = 0$ the condition $dQ(z)/dz = 0$ is easily
solvable. For this special
case one has to search for the minimum of function
\begin{equation}
\frac{(1+z)^{N+q-2}}{(1 - (1+z)^{-1/2})^2},
\end{equation}
because here $d_\mathrm{M} = (2c/H_0)/(1 - (1+z)^{-1/2})$. Then one has
\begin{equation}
z_\mathrm{turn} =\Bigl( \frac{{N+q-1}}{N+q-2}\Bigr)^2 - 1.
\end{equation}

\section{The samples}

\subsection{The choice of burst samples}

The frequency of the occurrence $\tilde{L}(z_1) < \tilde{L}(z_2)$
at $z_1 < z_2$, but for their observed values $P(z_1) <
P(z_2)$, i.e. the more distant GRB is apparently brighter for the observer, 
can be verified on a sample for which there are well-defined redshifts,
as well as measured peak-fluxes and/or fluences.
At a given redshift the luminosity $\tilde{L}(z)$ is a stochastic
variable and starting from  Eq.(\ref{eq:3})  one can get the probability for
$P'_{obs} > P(z)$, assuming that $f(\tilde{L}|z)$ is given.

There are two different subgroups of GRBs, which can be denoted as
"short-" and "long-"duration bursts (\cite{nor01,bal03,me06b,zha09}). In
addition, the existence of additional subgroups cannot be excluded
(\cite{ho98,ho02,hak03,bor04,var05,ho08,va08,ho09}). The first direct
redshift for a long (short) GRBs was determined in 1997 (2005)
(for a detailed survey see, e.g., Piran (2004) and M\'esz\'aros
(2006)). The few redshifts measured for short bursts are on average
small (\cite{me09}), which motivates
us to concentrate on long-duration bursts only in this study.

Since we try to obtain consequences of the GRBs' redshifts also in
the BATSE Catalog, we should obviously study the BATSE sample. But
only a few of these bursts have directly measured redshifts from
afterglow data. Therefore we will also try to obtain conclusions
from a sample that uses the so-called "pseudo-redshifts", i.e. the
redshifts estimated exclusively from the gamma photon measurements
alone. But for them one should keep in mind that they can be
uncertain. Thus, the best is to compare the BATSE samples with
other samples of long GRBs. The redshifts of GRBs detected by Swift
satellite - obtained from afterglow data - can well serve for this
comparison. On the other hand, the redshifts of GRBs - detected by
other satellites (BeppoSAX, HETE-2, INTEGRAL) - are not good for
our purpose, since they are strongly biased with selection
effects (\cite{lee00,bag06,bu10}), and represent only the brightest bursts.
All this means that we
will discuss four samples here: BATSE GRBs with known redshifts,
BATSE GRBs with pseudo-redshifts, the Swift sample and the Fermi sample.
We will try to show the occurrence of the inverse behaviour, first, without
the special assumptions of subsection 2.2., and, second,
using this subsection.

\subsection{Swift GRBs and the inversion in this sample}

In the period of 20 November 2004 $-$ 30 April
2010 the Swift database gives a sample of 134 bursts with well determined redshifts
from the afterglows together with BAT fluences and peak-fluxes in the energy
range of $15-150$\,keV. To
abandon the short bursts only those with $T_{90}/(1+z) > 2\,s$ were taken.
They are collected in Tables \ref{tab:SwiftI}-\ref{tab:SwiftIV}.

\begin{figure*}[]
\centering
\includegraphics[width=0.95\textwidth]{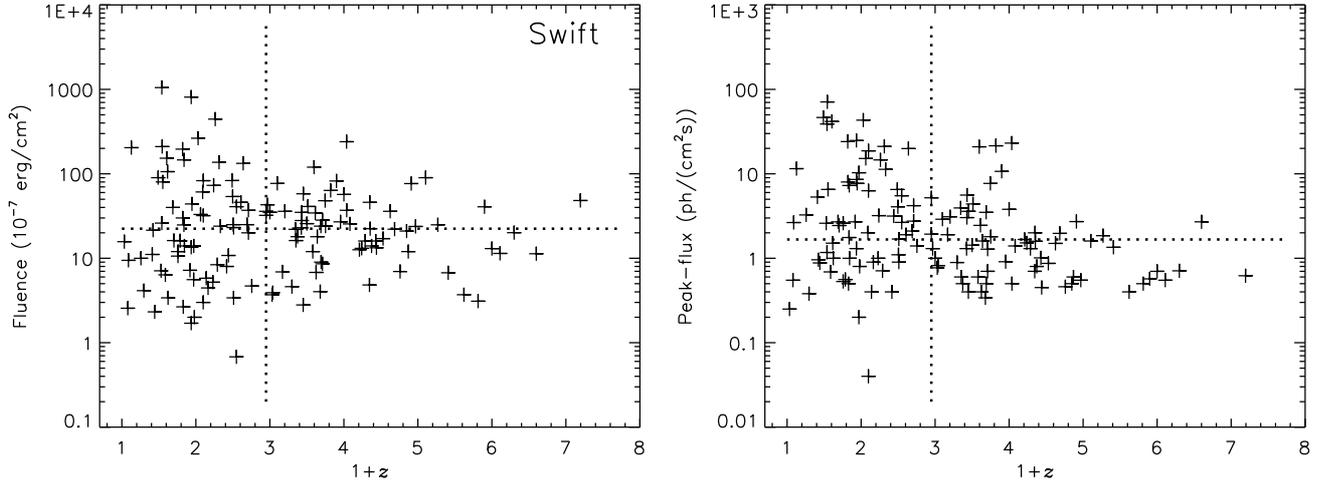}
\caption{Distribution of the fluences (left panel) and peak-fluxes (right panel) of the
Swift GRBs with known redshifts. The medians separate the area into four quadrants. The
objects in the upper right quadrant are brighter and have larger redshifts than the that
of GRBs in the lower left quadrant.}
\label{fig:swift}
\end{figure*}

The effect of inversion can be
demonstrated by the scatter plots of the [log fluence; $z$] and [log peak-flux; $z$] planes
as it can be seen in Fig. \ref{fig:swift}. To demonstrate the effect of inversion we marked the
medians of the
fluence and peak-flux with horizontal and that of the redshift with vertical dashed lines.
The medians split the plotting area into four quadrants. It is easy to see that GRBs in the
upper right quadrant are apparently brighter than those in the lower left one, although, their
redshifts are larger. It is worth mentioning that the GRB having the greatest redshift in the
sample has higher fluence than 50\% of all the items in the sample.

\begin{figure*}[]
\centering
\includegraphics[width=0.95\textwidth]{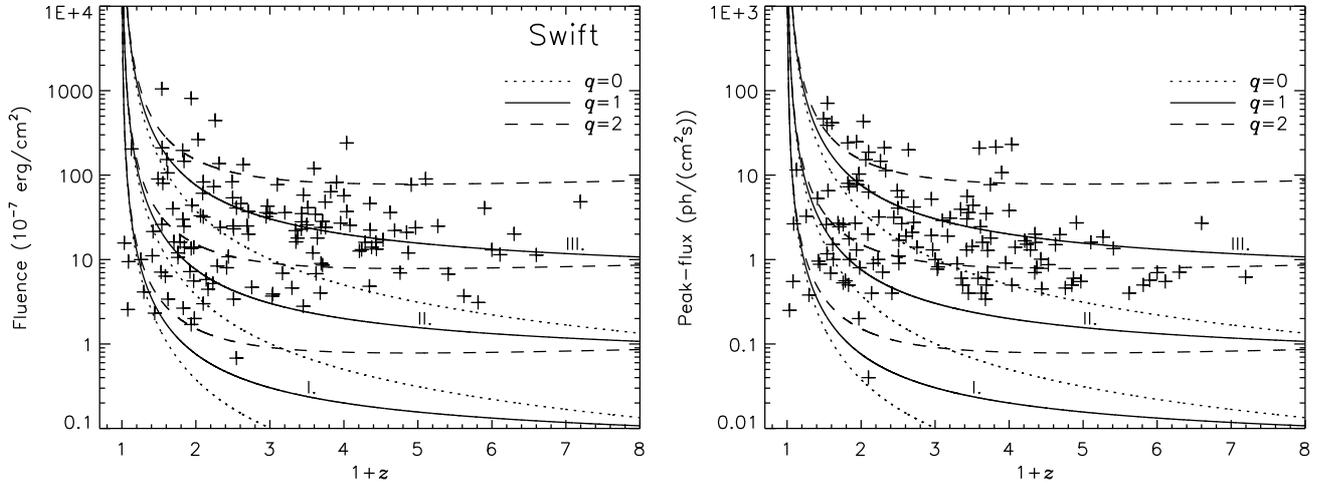}
\caption{Distribution of the fluences (left panel) and peak-fluxes (right panel) of the
Swift GRBs with known redshifts.
On the left panel the curves denote the values of fluences for $\tilde{E}_\mathrm{iso} = \tilde{E_o}
(1+z)^q $
(the three constant $\tilde{E_o}$ are in units $10^{51}$\,erg: I. 0.1; II. 1.0;
III.  10.0). On the right panel
the curves denote the values of peak-fluxes for $\tilde{L}_\mathrm{iso} =
\tilde{L_o} (1+z)^q$ (the three constant $\tilde{L_o}$ are in units $10^{58}$\,ph/s: I.
0.01; II. 0.1; III. 1.0).}
\label{fig:swiftflue}
\end{figure*}

Using the special assumption of subsection 2.2. the effect of inversion may be illustrated
in the Swift sample distinctly also as follows.
In Fig. \ref{fig:swiftflue} the fluences and peak-fluxes
are typified against the redshifts.
Be the luminosity distances calculated for $H_0 = 71$\,km/(s\,Mpc), $\Omega_\mathrm{M} =
0.27$ and $\Omega_\mathrm{\Lambda} = 0.73$.  Then the total emitted energy $\tilde{E}_\mathrm{iso}$
and the peak-luminosity $\tilde{L}_\mathrm{iso}$ can be calculated using Eq.(\ref{eq:2}) with
$N=1$. In the figure the curves of fluences and peak-fluxes are shown after substituting
$\tilde{L}_\mathrm{iso} =
\tilde{L_o} (1+z)^q$ and
$\tilde{E}_\mathrm{iso} = \tilde{E_o} (1+z)^q$ where $\tilde{L_o}$ and $\tilde{E_o}$ are
constants, and $q = 0; \;1, \;2$.
The inverse behaviour is quite obvious for $q >1$ and roughly for $z >2$.

\begin{figure*}[]
\centering
\includegraphics[width=0.95\textwidth]{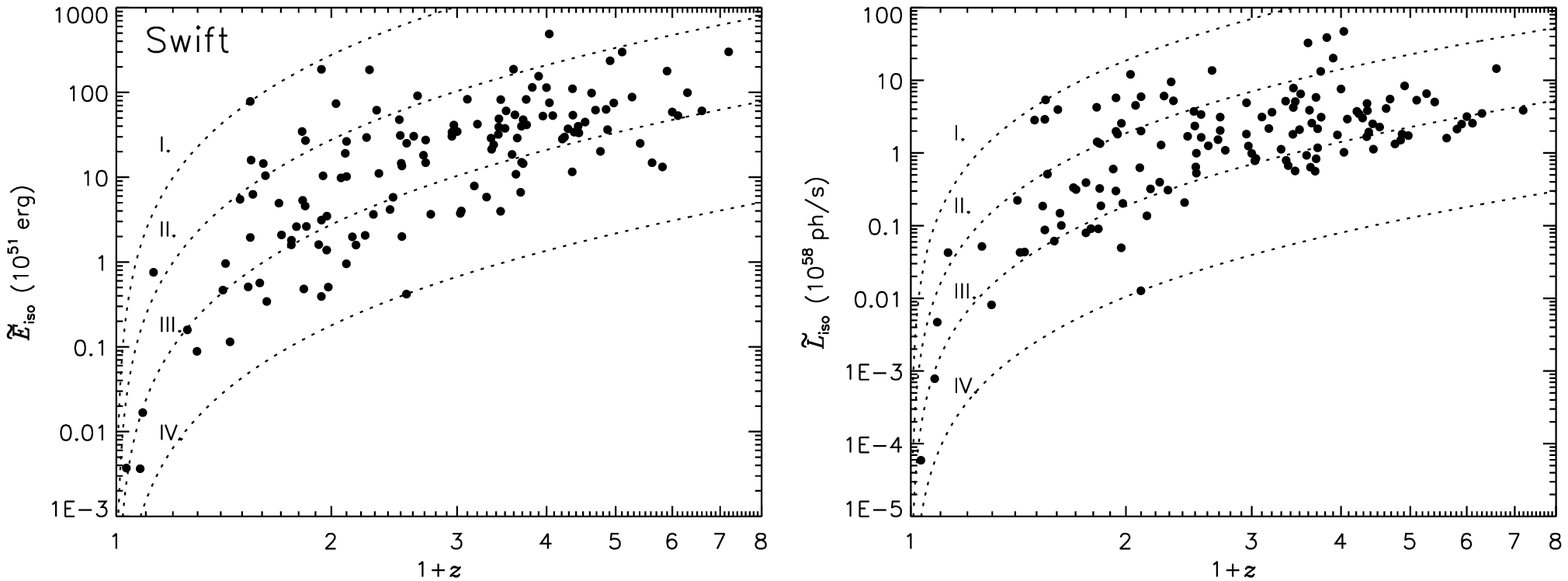}
\caption{\emph{Left panel}: $\tilde{E}_\mathrm{iso}$
vs. $(1+z)$ (dots). Dashed
contours denote constant fluences (in units $10^{-7}$\,erg/cm$^2$):
I. the maximal fluence, i.e. 1050; II. 105; III. 10.5; IV. the minimal fluence, i.e. 0.68.
 \emph{Right panel}: $\tilde{L}_\mathrm{iso}$ vs. $(1+z)$ (dots).
 Dashed contours denote constant peak-fluxes (in units ph\,cm$^{-2}$s$^{-1}$):
 I. the maximal peak-flux, i.e. 71; II. 7.1; III. 0.71; IV. the minimal peak-flux, i.e. 0.04.
The objects below a curve at smaller redshifts together with those at higher redshifts
and above the curve illustrate the inverse behaviour.}
\label{fig:swiftEiso}
\end{figure*}

The same effect can be similarly illustrated also in Fig. \ref{fig:swiftEiso} showing
the relation $ \tilde{E}_\mathrm{iso}$ vs. $(1+z)$,
and the relation $ \tilde{L}_\mathrm{iso}$ vs. $(1+z)$. They were calculated again for $H_0 =
71$\,km/(s\,Mpc), $\Omega_\mathrm{M} =
0.27$ and $\Omega_\mathrm{\Lambda} = 0.73$.
In the figure the curves of constant observable peak-fluxes and fluences are also shown.
These curves discriminate the bursts of lower/higher measured values. GRBs below a curve at
smaller redshifts are representing the inverse behaviour with respect to those at higher redshifts
and above the curve.

\begin{table*}
\begin{center}
\caption[]{Swift sample with known redshifts I.}
\begin{tabular}{lcccccc}
\hline
\\[-2.3ex]
GRB & Fluence   & Peak-flux    & $z$ & $d_\mathrm{l}$ & $\tilde{E}_\mathrm{iso}$ & $\tilde{L}_\mathrm{iso}$\\
    & $10^{-7}$\,erg/cm$^2$& ph/(cm$^2$s)& & Gpc & $10^{51}$\,erg  & $ 10^{58}$\,ph/s \\
 \hline
050126     &     8.38     &     0.71     &     1.29     &     9.12     &     3.6E+0     &     3.1E-1     \\
050223     &     6.36     &     0.69     &     0.588     &     3.44     &     5.7E-1     &     6.1E-2     \\
050315     &     32.2     &     1.93     &     1.949     &     15.24     &     3.0E+1     &     1.8E+0     \\
050318     &     10.8     &     3.16     &     1.44     &     10.46     &     5.8E+0     &     1.7E+0     \\
050319     &     13.1     &     1.52     &     3.24     &     28.37     &     3.0E+1     &     3.5E+0     \\
050401     &     82.2     &     10.7     &     2.9     &     24.81     &     1.6E+2     &     2.0E+1     \\
050505     &     24.9     &     1.85     &     4.27     &     39.49     &     8.8E+1     &     6.6E+0     \\
050525A     &     153     &     41.7     &     0.606     &     3.57     &     1.5E+1     &     4.0E+0     \\
050603     &     63.6     &     21.5     &     2.821     &     23.99     &     1.1E+2     &     3.9E+1     \\
050724     &     9.98     &     3.26     &     0.258     &     1.29     &     1.6E-1     &     5.2E-2     \\
050730     &     23.8     &     0.55     &     3.969     &     36.20     &     7.5E+1     &     1.7E+0     \\
050802     &     20     &     2.75     &     1.71     &     12.96     &     1.5E+1     &     2.0E+0     \\
050803     &     21.5     &     0.96     &     0.422     &     2.30     &     9.6E-1     &     4.3E-2     \\
050814     &     20.1     &     0.71     &     5.3     &     50.99     &     9.9E+1     &     3.5E+0     \\
050820A     &     34.4     &     2.45     &     2.613     &     21.85     &     5.4E+1     &     3.9E+0     \\
050824     &     2.66     &     0.5     &     0.83     &     5.26     &     4.8E-1     &     9.0E-2     \\
050826     &     4.13     &     0.38     &     0.297     &     1.52     &     8.8E-2     &     8.1E-3     \\
050904     &     48.3     &     0.62     &     6.195     &     61.22     &     3.0E+2     &     3.9E+0     \\
050908     &     4.83     &     0.7     &     3.346     &     29.49     &     1.2E+1     &     1.7E+0     \\
051016B     &     1.7     &     1.3     &     0.936     &     6.11     &     3.9E-1     &     3.0E-1     \\
051109A     &     22     &     3.94     &     2.346     &     19.15     &     2.9E+1     &     5.2E+0     \\
051109B     &     2.56     &     0.55     &     0.08     &     0.36     &     3.7E-3     &     7.8E-4     \\
051111     &     40.8     &     2.66     &     1.549     &     11.46     &     2.5E+1     &     1.6E+0     \\
060108     &     3.69     &     0.77     &     2.03     &     16.03     &     3.7E+0     &     7.8E-1     \\
060115     &     17.1     &     0.87     &     3.53     &     31.45     &     4.5E+1     &     2.3E+0     \\
060123     &     3     &     0.04     &     1.099     &     7.47     &     9.5E-1     &     1.3E-2     \\
060124     &     4.61     &     0.89     &     2.298     &     18.67     &     5.8E+0     &     1.1E+0     \\
060210     &     76.6     &     2.72     &     3.91     &     35.55     &     2.4E+2     &     8.4E+0     \\
060218     &     15.7     &     0.25     &     0.033     &     0.14     &     3.7E-3     &     5.9E-5     \\
060223A     &     6.73     &     1.35     &     4.41     &     41.03     &     2.5E+1     &     5.0E+0     \\
060418     &     83.3     &     6.52     &     1.49     &     10.91     &     4.8E+1     &     3.7E+0     \\
\hline
\end{tabular}
\label{tab:SwiftI}
\end{center}
\end{table*}

\begin{table*}
\begin{center}
\caption[]{Swift sample with known redshifts II.}
\begin{tabular}{lcccccc}
\hline
\\[-2.3ex]
GRB & Fluence   & Peak-flux    & $z$ & $d_\mathrm{l}$ & $\tilde{E}_\mathrm{iso}$ & $\tilde{L}_\mathrm{iso}$\\
    & $10^{-7}$\,erg/cm$^2$& ph/(cm$^2$s)& & Gpc & $10^{51}$\,erg  & $ 10^{58}$\,ph/s \\
 \hline
060502A     &     23.1     &     1.69     &     1.51     &     11.10     &     1.4E+1     &     9.9E-1     \\
060505     &     9.44     &     2.65     &     0.089     &     0.40     &     1.7E-2     &     4.7E-3     \\
060510B     &     40.7     &     0.57     &     4.9     &     46.49     &     1.8E+2     &     2.5E+0     \\
060512     &     2.32     &     0.88     &     0.443     &     2.44     &     1.1E-1     &     4.3E-2     \\
060522     &     11.4     &     0.55     &     5.11     &     48.85     &     5.3E+1     &     2.6E+0     \\
060526     &     12.6     &     1.67     &     3.21     &     28.05     &     2.8E+1     &     3.7E+0     \\
060602A     &     16.1     &     0.56     &     0.787     &     4.92     &     2.6E+0     &     9.1E-2     \\
060604     &     4.02     &     0.34     &     2.68     &     22.54     &     6.6E+0     &     5.6E-1     \\
060605     &     6.97     &     0.46     &     3.76     &     33.93     &     2.0E+1     &     1.3E+0     \\
060607A     &     25.5     &     1.4     &     3.082     &     26.71     &     5.3E+1     &     2.9E+0     \\
060614     &     204     &     11.5     &     0.128     &     0.59     &     7.6E-1     &     4.3E-2     \\
060707     &     16     &     1.01     &     3.43     &     30.39     &     4.0E+1     &     2.5E+0     \\
060714     &     28.3     &     1.28     &     2.71     &     22.84     &     4.8E+1     &     2.2E+0     \\
060729     &     26.1     &     1.17     &     0.54     &     3.10     &     1.9E+0     &     8.7E-2     \\
060814     &     146     &     7.27     &     0.84     &     5.34     &     2.7E+1     &     1.3E+0     \\
060904B     &     16.2     &     2.44     &     0.703     &     4.28     &     2.1E+0     &     3.1E-1     \\
060906     &     22.1     &     1.97     &     3.685     &     33.12     &     6.2E+1     &     5.5E+0     \\
060908     &     28     &     3.03     &     2.43     &     20.00     &     3.9E+1     &     4.2E+0     \\
060912     &     13.5     &     8.58     &     0.937     &     6.12     &     3.1E+0     &     2.0E+0     \\
060927     &     11.3     &     2.7     &     5.6     &     54.40     &     6.1E+1     &     1.4E+1     \\
061007     &     444     &     14.6     &     1.262     &     8.87     &     1.8E+2     &     6.1E+0     \\
061110A     &     10.6     &     0.53     &     0.758     &     4.70     &     1.6E+0     &     8.0E-2     \\
061110B     &     13.3     &     0.45     &     3.44     &     30.49     &     3.3E+1     &     1.1E+0     \\
061121     &     137     &     21.1     &     1.314     &     9.33     &     6.2E+1     &     9.5E+0     \\
061210     &     11.1     &     5.31     &     0.41     &     2.22     &     4.7E-1     &     2.2E-1     \\
061222B     &     22.4     &     1.59     &     3.355     &     29.59     &     5.4E+1     &     3.8E+0     \\
070110     &     16.2     &     0.6     &     2.352     &     19.21     &     2.1E+1     &     7.9E-1     \\
070208     &     4.45     &     0.9     &     1.165     &     8.03     &     1.6E+0     &     3.2E-1     \\
070306     &     53.8     &     4.07     &     1.497     &     10.98     &     3.1E+1     &     2.4E+0     \\
070318     &     24.8     &     1.76     &     0.838     &     5.32     &     4.6E+0     &     3.2E-1     \\
070411     &     27     &     0.91     &     2.954     &     25.37     &     5.3E+1     &     1.8E+0     \\
070419A     &     5.58     &     0.2     &     0.97     &     6.39     &     1.4E+0     &     5.0E-2     \\
070508     &     196     &     24.1     &     0.82     &     5.18     &     3.5E+1     &     4.3E+0     \\
070521     &     80.1     &     6.53     &     0.553     &     3.19     &     6.3E+0     &     5.1E-1     \\
070529     &     25.7     &     1.43     &     2.5     &     20.70     &     3.8E+1     &     2.1E+0     \\
070611     &     3.91     &     0.82     &     2.04     &     16.13     &     4.0E+0     &     8.4E-1     \\
070612A     &     106     &     1.51     &     0.617     &     3.65     &     1.0E+1     &     1.5E-1     \\
\hline
\end{tabular}
\label{tab:SwiftII}
\end{center}
\end{table*}

\begin{table*}
\begin{center}
\caption[]{Swift sample with known redshifts III.}
\begin{tabular}{lcccccc}
\hline
\\[-2.3ex]
GRB & Fluence   & Peak-flux    & $z$ & $d_\mathrm{l}$ & $\tilde{E}_\mathrm{iso}$ & $\tilde{L}_\mathrm{iso}$\\
    & $10^{-7}$\,erg/cm$^2$& ph/(cm$^2$s)& & Gpc & $10^{51}$\,erg  & $ 10^{58}$\,ph/s \\
 \hline
070714B     &     7.2     &     2.7     &     0.92     &     5.98     &     1.6E+0     &     6.0E-1     \\
070721B     &     36     &     1.5     &     3.626     &     32.48     &     9.8E+1     &     4.1E+0     \\
070802     &     2.8     &     0.4     &     2.45     &     20.20     &     4.0E+0     &     5.7E-1     \\
070810A     &     6.9     &     1.9     &     2.17     &     17.40     &     7.9E+0     &     2.2E+0     \\
071003     &     83     &     6.3     &     1.1     &     7.47     &     2.6E+1     &     2.0E+0     \\
071010A     &     2     &     0.8     &     0.98     &     6.47     &     5.1E-1     &     2.0E-1     \\
071010B	    &	44	&	7.7    & 0.947	&	6.20	    &	  1.0E+1	 &     	1.8E+0	 \\
071021     &     13     &     0.7     &     5.0     &     47.61     &     5.9E+1     &     3.2E+0     \\
071031     &     9     &     0.5     &     2.692     &     22.66     &     1.5E+1     &     8.3E-1     \\
071112C     &     30     &     8     &     0.823     &     5.20     &     5.3E+0     &     1.4E+0     \\
071117     &     24     &     11.3     &     1.331     &     9.48     &     1.1E+1     &     5.2E+0     \\
071122     &     5.8     &     0.4     &     1.14     &     7.82     &     2.0E+0     &     1.4E-1     \\
080210     &     18     &     1.6     &     2.641     &     22.13     &     2.9E+1     &     2.6E+0     \\
080310     &     23     &     1.3     &     2.426     &     19.95     &     3.2E+1     &     1.8E+0     \\
080319B    &	810	 &	24.8     &  	0.937	 &	    6.12	  &  	1.9E+2	   &	5.7E+0	 \\
080319C     &     36     &     5.2     &     1.95     &     15.25     &     3.4E+1     &     4.9E+0     \\
080330     &     3.4     &     0.9     &     1.51     &     11.10     &     2.0E+0     &     5.3E-1     \\
080411     &     264     &     43.2     &     1.03     &     6.88     &     7.4E+1     &     1.2E+1     \\
080413A     &     35     &     5.6     &     2.432     &     20.01     &     4.9E+1     &     7.8E+0     \\
080413B     &     32     &     18.7     &     1.1     &     7.47     &     1.0E+1     &     6.0E+0     \\
080430     &     12     &     2.6     &     0.759     &     4.70     &     1.8E+0     &     3.9E-1     \\
080603B     &     24     &     3.5     &     2.69     &     22.64     &     4.0E+1     &     5.8E+0     \\
080604     &     8     &     0.4     &     1.416     &     10.25     &     4.2E+0     &     2.1E-1     \\
080605     &     133     &     19.9     &     1.64     &     12.30     &     9.1E+1     &     1.4E+1     \\
080607     &     240     &     23.1     &     3.036     &     26.22     &     4.9E+2     &     4.7E+1     \\
080707     &     5.2     &     1     &     1.23     &     8.59     &     2.1E+0     &     4.0E-1     \\
080710     &     14     &     1     &     0.845     &     5.38     &     2.6E+0     &     1.9E-1     \\
080721     &     120     &     20.9     &     2.597     &     21.68     &     1.9E+2     &     3.3E+1     \\
080804     &     36     &     3.1     &     2.202     &     17.72     &     4.2E+1     &     3.6E+0     \\
080805     &     25     &     1.1     &     1.505     &     11.06     &     1.5E+1     &     6.4E-1     \\
080810     &     46     &     2     &     3.35     &     29.53     &     1.1E+2     &     4.8E+0     \\
080905B     &     18     &     0.5     &     2.374     &     19.43     &     2.4E+1     &     6.7E-1     \\
080906     &     35     &     1     &     2     &     15.74     &     3.5E+1     &     9.9E-1     \\
080916A     &     40     &     2.7     &     0.689     &     4.18     &     4.9E+0     &     3.3E-1     \\
080928     &     25     &     2.1     &     1.691     &     12.78     &     1.8E+1     &     1.5E+0     \\
\hline
\end{tabular}
\label{tab:SwiftIII}
\end{center}
\end{table*}

\begin{table*}
\begin{center}
\caption[]{Swift sample with known redshifts IV.}
\begin{tabular}{lcccccc}
\hline
\\[-2.3ex]
GRB & Fluence   & Peak-flux    & $z$ & $d_\mathrm{l}$ & $\tilde{E}_\mathrm{iso}$ & $\tilde{L}_\mathrm{iso}$\\
    & $10^{-7}$\,erg/cm$^2$& ph/(cm$^2$s)& & Gpc & $10^{51}$\,erg  & $ 10^{58}$\,ph/s \\
 \hline
081007     &     7.1     &     2.6     &     0.53     &     3.03     &     5.1E-1     &     1.9E-1     \\
081008     &     43     &     1.3     &     1.968     &     15.42     &     4.1E+1     &     1.2E+0     \\
081028A     &     37     &     0.5     &     3.038     &     26.24     &     7.6E+1     &     1.0E+0     \\
081029     &     21     &     0.5     &     3.847     &     34.88     &     6.3E+1     &     1.5E+0     \\
081118     &     12     &     0.6     &     2.58     &     21.51     &     1.9E+1     &     9.3E-1     \\
081121     &     41     &     4.4     &     2.512     &     20.82     &     6.1E+1     &     6.5E+0     \\
081203A     &     77     &     2.9     &     2.1     &     16.71     &     8.3E+1     &     3.1E+0     \\
081222     &     48     &     7.7     &     2.747     &     23.22     &     8.3E+1     &     1.3E+1     \\
090102     &     0.68     &     5.5     &     1.548     &     11.45     &     4.2E-1     &     3.4E+0     \\
090313     &     14     &     0.8     &     3.374     &     29.79     &     3.4E+1     &     1.9E+0     \\
090418A     &     46     &     1.9     &     1.608     &     12.01     &     3.0E+1     &     1.3E+0     \\
090424     &     210     &     71     &     0.544     &     3.13     &     1.6E+1     &     5.4E+0     \\
090516A     &     90     &     1.6     &     4.105     &     37.68     &     3.0E+2     &     5.3E+0     \\
090519     &     12     &     0.6     &     3.87     &     35.12     &     3.6E+1     &     1.8E+0     \\
090529   	&	6.8	    &	  0.4	  &	   2.62 &   21.97 &	   1.1E+1	  &	6.4E-1	 \\
090618     &     1050     &     38.9     &     0.54     &     3.10     &     7.8E+1     &     2.9E+0     \\
090715B     &     57     &     3.8     &     3     &     25.85     &     1.1E+2     &     7.6E+0     \\
090726     &     8.6     &     0.7     &     2.71     &     22.84     &     1.4E+1     &     1.2E+0     \\
090812     &     58     &     3.6     &     2.452     &     20.22     &     8.2E+1     &     5.1E+0     \\
090926B     &     73     &     3.2     &     1.24     &     8.68     &     2.9E+1     &     1.3E+0     \\
091018     &     14     &     10.3     &     0.971     &     6.40     &     3.5E+0     &     2.6E+0     \\
091020     &     37     &     4.2     &     1.71     &     12.96     &     2.7E+1     &     3.1E+0     \\
091024     &     61     &     2     &     1.092     &     7.40     &     1.9E+1     &     6.3E-1     \\
091029     &     24     &     1.8     &     2.752     &     23.28     &     4.1E+1     &     3.1E+0  \\
091109A	   &	16	   &	1.3	    &	3.288	   &	28.88	   &	3.7E+1	   &	3.0E+0	\\
091127	   &	90	   &	46.5	&	0.49	   &	2.75	   &	5.5E+0	   &	2.8E+0	\\
091208B	   &	33	   &	15.2	&	1.063	   &	7.16	   &	9.8E+0	   &	4.5E+0	\\
100219A	   &	3.7	   &	0.4	    &	4.622	   &	43.38	   &	1.5E+1	   &	1.6E+0	\\
100302A	   &	3.1    &	0.5	    &	4.813	   &	45.51	   &	1.3E+1	   &	2.1E+0	\\
100418A	   &	3.4	   &	1	    &	0.624	   &	3.70	   &	3.4E-1	   &	1.0E-1	\\
100425A	   &	4.7	   &	1.4	    &	1.755	   &	13.38	   &	3.7E+0	   &	1.1E+0	\\

\hline
\end{tabular}
\label{tab:SwiftIV}
\end{center}
\end{table*}

\subsection{BATSE sample with known redshifts}

There are 11 bursts, which were observed by BATSE during the
period 1997-2000 and for which there are observed redshifts from
the afterglow data.
For one of them, GRB980329 (BATSE trigger 6665), the redshift has
only an upper limit ($z < 3.5$), and hence will not be used here.
Two cases (GRB970828 [6350] $z$ = 0.9578 and GRB000131 [7975] $z$
= 4.5) have determined redshifts, but they have no measured
fluences and peak-fluxes, hence they are also excluded.
There are remaining 8 GRBs, which are collected in Table \ref{tab:BATSEz}
(see also Bagoly et al. (2003) and Borgonovo (2004)).
For the definition of fluence we have
chosen the fluence from the third BATSE channel between 100 and
300\,keV ($F_3$). This choice is motivated by the observational
fact that these fluences in the BATSE Catalog are usually well
measured and correlate with other fluences (\cite{bag98,ver05}).

\begin{table*}
\begin{center}
\caption[]{BATSE sample with known redshifts.}
\begin{tabular}{cccccccc}

\hline
\\[-2.3ex]
GRB & BATSE  &   $z$   &   $F_3$ & $P_{256}$ & $d_\mathrm{l}$ & $\tilde{E}_\mathrm{iso}$ &  $\tilde{L}_\mathrm{iso}$ \\
&trigger & & $10^{-6}$\,erg/cm$^2$ & ph/(cm$^2$s) & Gpc & $10^{51}$\,erg &$10^{58}$\,ph/s
\\
\hline
970508 & 6225 &    0.835 &    0.88 &      1.18&    5.3 &    1.6E+0 &    2.2E-1  \\
971214 & 6533 &    3.42  &    4.96 &      2.32&   30.3 &    1.2E+2 &    5.8E+0  \\
980425 & 6707 &    0.0085&    1.67 &      1.08&    0.036 &  2.6E-4 &    1.7E-5  \\
980703 & 6891 &    0.966 &   14.6  &      2.59&    6.35 &   3.6E+1 &    6.4E-1  \\
990123 & 7343 &    1.600 &   87.2  &     16.63&   11.9 &    5.7E+2 &    1.1E+1  \\
990506 & 7549 &    1.307 &   51.6  &     22.16&    9.3 &    2.3E+2 &    9.9E+0  \\
990510 & 7560 &    1.619 &    8.0  &     10.19&   12.1 &    5.4E+1 &    6.8E+0  \\
991216 & 7906 &    1.02  &   63.7  &     82.10&    6.8 &    1.7E+2 &    2.3E+1  \\
\hline
\end{tabular}
\label{tab:BATSEz}
\end{center}
\end{table*}

\subsection{BATSE pseudo-redshifts}

In the choice of a BATSE sample with estimated pseudo-redshifts
one has to take care, since these redshifts are less reliable than
the direct redshifts from AGs. We will use here the
pseudo-redshifts based on the luminosity-lag relation, restricted
to the sample in Ryde et al. (2005), where also the spectroscopic
studies support the estimations.
In Table \ref{tab:BATSEpseu} we collect 13 GRBs using Table 3 of Ryde et al. (2005).
We do not consider two GRBs (BATSE triggers 973 and 3648) from Ryde et al.
(2005), since their estimated pseudo-redshifts are ambiguous.

\begin{table*}
\begin{center}
\caption[]{BATSE sample with pseudo-redshifts.}
\begin{tabular}{cccccccc}
\hline
\\[-2.3ex]
GRB & BATSE & $z$ &   $F_3$     & $P_{256}$    & $d_\mathrm{l}$ & $\tilde{E}_\mathrm{iso}$ & $\tilde{L}_\mathrm{iso}$\\
    & trigger &   & $10^{-6}$\,erg/cm$^2$ & ph/(cm$^2$s) & Gpc & $10^{51}$\,erg & $10^{58}$\,ph/s\\
 \hline
911016  & 907  &  0.40 & 8.2   & 3.6   & 2.2  & 3.4E+0 &  1.5E-1 \\
911104  & 999  &  0.67 & 2.6   & 11.5  & 4.1  & 3.1E+0 &  1.4E+0 \\
920525  & 1625 &  1.8  & 27.2  & 27.3  & 13.8 & 2.2E+2 &  2.3E+1 \\
920830  & 1883 &  0.45 & 3.5   & 5.2   &  2.5 & 1.8E+0 &  2.7E-1 \\
921207  & 2083 &  0.18 & 30.5  & 45.4  & 0.86 & 2.3E+0 &  3.4E-1 \\
930201  & 2156 &  0.41 & 59.6  & 16.6  & 2.2  & 2.5E+1 &  6.6E-1 \\
941026  & 3257 &  0.38 & 8.4   & 3.1   &  2.1 & 3.2E+0 &  1.2E-1 \\
950818  & 3765 &  0.64 & 19.2  & 25.3  & 3.8  & 2.0E+1 &  2.7E+0 \\
951102  & 3891 &  0.68 & 4.8   & 13.7  &  4.1 & 5.8E+0 &  1.6E+0 \\
960530  & 5478 & 0.53  & 5.9   &  3.0  &  3.0 & 4.2E+0 &  2.1E-1 \\
960804  & 5563 & 0.76  & 2.7   & 21.4  &  4.7 & 4.1E+0 &  3.2E+0 \\
980125  & 6581 & 1.16  & 16.3  & 38.4  &  8.0 & 5.8E+1 &  1.4E+1 \\
990102  & 7293 & 8.6   &  6.3  &  2.9  & 89.4 & 6.3E+2 &  2.8E+1 \\
\hline
\end{tabular}
\label{tab:BATSEpseu}
\end{center}
\end{table*}

\subsection{Fermi sample}

The Fermi sample contains only 6 GRBs with known redshifts together with
peak-fluxes and fluences (\cite{bi09a,bi09b,horst09,ra09a,ra09b,fo10}). They
are collected in Table \ref{tab:Fermi}. The peak-fluxes and fluences were measured
over the energy range $50\,\mathrm{keV} - 10\,\mathrm{MeV}$ for GRB090902B and in the
range $8\,\mathrm{keV} - 1\,\mathrm{MeV}$ for the remaining five objects.

\begin{table*}
\begin{center}
\caption[]{Fermi sample with known redshifts.}
\begin{tabular}{lcccccc}
\hline
\\[-2.3ex]
GRB & Fluence   & Peak-flux    & $z$ & $d_\mathrm{l}$ & $\tilde{E}_\mathrm{iso}$ & $\tilde{L}_\mathrm{iso}$\\
    & $10^{-7}$\,erg/cm$^2$& ph/(cm$^2$s)&  & Gpc & $10^{51}$\,erg  & $ 10^{58}$\,ph/s \\
 \hline
090323	&	1000	&	12.3	&	3.79	&	31.88	&	2.66E+3	&	3.27E+1	\\
090328	&	809	    &	18.5	&	0.736	&	4.53	&	1.34E+2	&	2.62E+0	\\
090902B	&	3740	&	46.1	&	1.822	&	14.02	&	3.12E+3	&	3.84E+1	\\
090926A	&	1450	&	80.8	&	2.106	&	16.77	&	1.57E+3	&	8.76E+1	\\
091003	&	376	    &	31.8	&	0.897	&	5.79	&	7.96E+1	&	6.73E+0	\\
100414A	&	1290	&	18.22	&	1.368	&	9.81	&	6.28E+2	&	8.87E+0	\\
\hline
\end{tabular}
\label{tab:Fermi}
\end{center}
\end{table*}

\subsection{Inversion in the BATSE and Fermi samples}

\begin{figure*}[]
\centering
\includegraphics[width=0.95\textwidth]{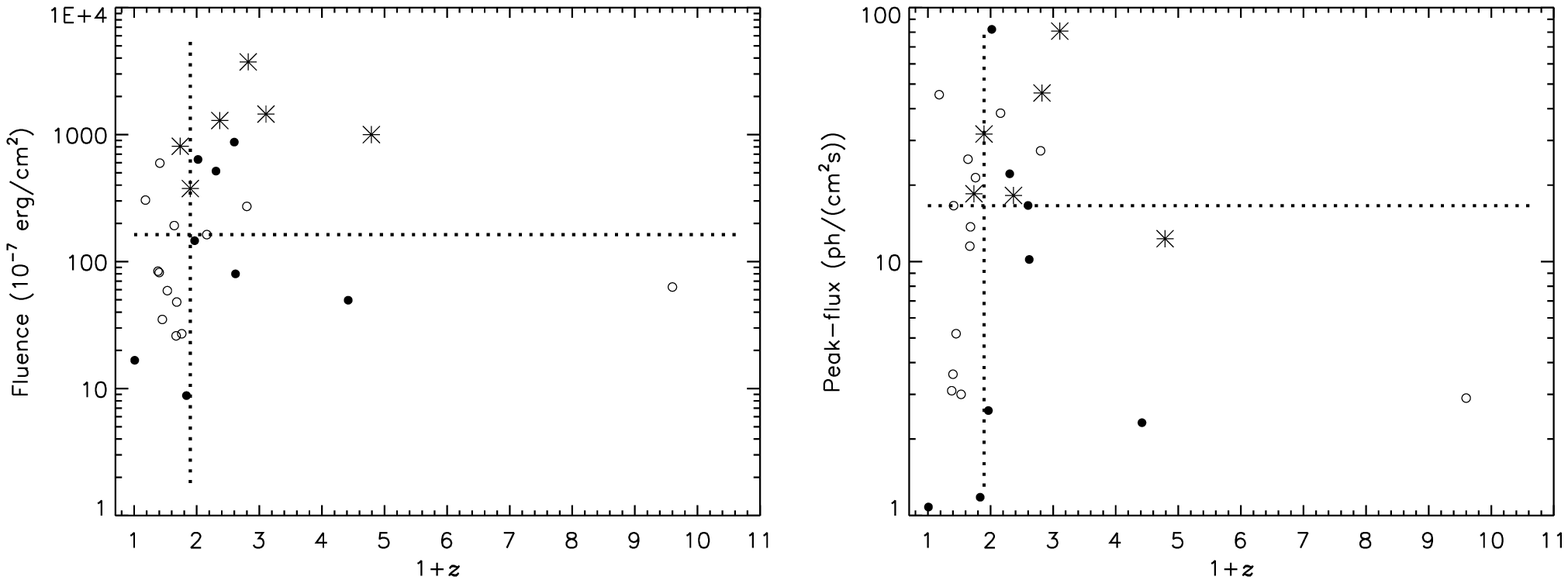}
\caption{Distribution of the fluences (left panel) and peak-fluxes (right panel) of the
GRBs with known redshifts, where the Fermi GRBs are denoted by asterisk, BATSE GRBs with
determined redshifts (pseudo-redshifts) are denoted by dots (circles). The medians separate the
area into four quadrants. The
objects in the upper right quadrant are brighter and have larger redshifts than the that
of GRBs in the lower left quadrant.}
\label{fig:allmedians}
\end{figure*}

The fluence (peak-flux) vs. redshift relationship of the Fermi and of the two
BATSE samples are summarized in Fig. \ref{fig:allmedians}. For demonstrating the inversion effect - 
similarly to the case of the Swift sample - the medians also marked with dashed lines. Here it is
quite evident that some of the distant bursts exceed in their observed fluence
and peak-fluxes those of having smaller redshifts. Here again the GRBs in the
upper right quadrants are apparently brighter than those in the lower left one, although their
redshifts are larger. Note that in the upper right quadrants are even more populated than the
lower right quadrants. In other words, here the trend of the increasing of peak-flux (fluence)
with redshift is evident, and the assumption of the subsection 2.2. need not be used.

\section{Results and discussion}

It follows from the previous section that in all samples both for the fluences and
for the peak-fluxes
the "inverse" behaviour, discussed in Section 2, can happen.
The answer on the question of the title of this article is therefore that
"this does not need to be the case." Simply, the apparently faintest GRBs need not be
also the most distant ones. This is in essence the key result of
this article.

It is essential to note that in this paper no assumptions
were made about the models of the long GRBs. Also the cosmological parameters
did not need to be specified.

All this means that faint bursts in the BATSE Catalog simply need
not be at larger redshifts, because the key "natural" assumption -
apparently fainter GRBs should on average be at higher redshifts - does
not hold. All this
also means that the controversy about the fraction of GRBs at very
high redshifts in BATSE Catalog may well be overcame: No large
fraction of GRBs needs to be at very high redshifts, and the
redshift distribution of long GRBs - coming from the Swift sample
- may well occur also for the BATSE sample. Of course, this does
not mean that no GRBs can be at, say, $8.3 < z < 20$. As proposed
first by M\'esz\'aros \& M\'esz\'aros (1996), at these very high
redshifts GRBs may well occur, but should give a minority (say
10 \% or so) in the BATSE Catalog similarly the Swift sample. This point
of view is supported by newer studies, too (Bromm \& Loeb 2006,
Jakobsson et al. 2006, Gehrels et al. 2009, Butler et al. 2010).

At the end it is useful to raise again that the purpose of this paper was not to study the
intrinsic evolution of the luminosities $L(z)$ from the energy range $E_1 \leq E \leq E_2$. To carry
out such study, one should consider three additional reasons
that may be responsible for the growth of the average value of
$\tilde{L}_\mathrm{iso}$ with $(1+z)$:
1. K-correction;
2. the beaming;
3. selection biases due to the instrument's threshold and other instrumental effects.
For example, on Fig.4 the main parts in the right-bottom sections below the curves IV -
corresponding to the values below the instrumental thresholds in fluence/peak-flux - are not observable.
Nonetheless, even using these biased data, the theoretical considerations stated in the
Section 2 and conclusions of the next Sections are valid.

\section{Conclusions}

The results of this paper can be summarized as follows:

1. The theoretical study of the $z$-dependence of the observed
 fluences and peak-fluxes of GRBs
have shown that fainter bursts could well have smaller redshifts.

2. This is really fulfilled for the four different samples of long GRBs.

3. These results do not depend on the cosmological parameters and on the GRB models.

4. All this suggests that the estimations, leading to a large
fraction of BATSE bursts at $z > 5$, need not be correct.

\begin{acknowledgements}
We wish to thank Z. Bagoly, L.G. Bal\'azs, I. Horv\'ath, S. Larsson,
P. M\'esz\'aros and P. Veres for useful discussions and comments on the
manuscript. The useful remarks of the anonymous referee are kindly acknow\-ledged.
This study was supported by the OTKA
grant K77795, by the Grant Agency of the Czech Republic grants No.
205/08/H005, and P209/10/0734, by the project SVV 261301 of the Charles
University in Prague, by the Research Program MSM0021620860 of the Ministry
of Education of the Czech Republic, and by the Swedish National Space Agency.

\end{acknowledgements}


\begin{thebibliography}{}

\bibitem[Amati et al. 2002]{am02} Amati, L., et al. 2002, A\&A, 390, 81

\bibitem[Atteia 2003]{atte03} Atteia, J-L. 2003, A\&A, 407, L1

\bibitem[Bagoly et al. 1998]{bag98} Bagoly, Z., et al. 1998, ApJ, 498, 342

\bibitem[Bagoly et al. 2003]{bag03} Bagoly, Z., et al. 2003, A\&A, 398, 919

\bibitem[Bagoly et al. 2006]{bag06} Bagoly, Z. et al. 2006, A\&A,
453, 797

\bibitem[Bagoly \& Veres 2009a]{bv09a} Bagoly, Z. \& Veres, P. 2009a,
Gamma-ray burst: Sixth Huntsville Symp., AIPC 1133, 473

\bibitem[Bagoly \& Veres 2009b]{bv09b} Bagoly, Z. \& Veres, P. 2009b,
Baltic Astronomy, 18, 297

\bibitem[Bal\'azs et al. 2003]{bal03} Bal\'azs, L.G., et al. 2003, A\&A, 401, 129

\bibitem[Band et al. 2004]{band04} Band, D. et al. 2004, ApJ, 613, 484

\bibitem[Bissaldi et al. 2009a]{bi09a} Bissaldi, E. et al. 2009a, GCN,
9866

\bibitem[Bissaldi et al. 2009b]{bi09b} Bissaldi, E. et al. 2009b, GCN,
9933

\bibitem[Borgonovo 2004]{bor04} Borgonovo, L. 2004, A\&A, 418, 487

\bibitem[Bromm \& Loeb 2006]{brolo06} Bromm, V. \& Loeb, A. 2006,
ApJ, 642, 382

\bibitem[Butler et al. 2010]{bu10} Butler, N.R., Bloom, J.S. \&
Poznanski, D. 2010, ApJ, 711, 495

\bibitem[Carroll et al. 1992]{ca92} Carroll, S.M., Press, W.H. \& Turner, E.L.
1992, ARA\&A, 30, 499

\bibitem[Foley et al. 2010]{fo10} Foley, S. et al. 2010, GCN, 10595

\bibitem[Gehrels et al. 2009]{ge09} Gehrels, N., et al. 2009,
ARA\&A, 47, 567

\bibitem[Ghirlanda et al. 2005]{ghi05} Ghirlanda, G., et al. 2005, MNRAS, 361, L10

\bibitem[Guidorzi et al. (2005)]{gui05} Guidorzi, C., et al. 2005, MNRAS, 363, 315

\bibitem[Hakkila et al. 2003]{hak03} Hakkila, J., et al. 2003, ApJ, 582, 320

\bibitem[Horv\'ath 1998]{ho98} Horv\'ath, I. 1998, ApJ, 508, 757

\bibitem[Horv\'ath 2002]{ho02} Horv\'ath, I. 2002, A\&A, 392, 791

\bibitem[Horv\'ath et al. 1996]{ho96} Horv\'ath, I.,
M\'esz\'aros, P. \& M\'esz\'aros, A. 1996, ApJ, 470, 56

\bibitem[Horv\'ath et al. 2008]{ho08} Horv\'ath, I.,
et al. 2008, A\&A, 489, L1

\bibitem[Horv\'ath et al. 2009]{ho09} Horv\'ath, I.,
et al. 2009, Fermi Symposium, eConf Proc. C0911022; astro-ph/0912.3724

\bibitem[Jakobsson et al. 2006]{jak06} Jakobsson, P., et al. 2006,
A\&A, 447, 897

\bibitem[Lamb \& Reichart 2000]{lare00}
 Lamb, D.Q. \& Reichart, D.E. 2000, ApJ, 536, 1

\bibitem[Lee et al. 2000]{lee00} Lee, A., Bloom, E.D. \&
Petrosian, V. 2000, ApJS, 131, 21

\bibitem[Lin et al. 2004]{lin04} Lin, J.R., Zhang, S.N. \&
Li, T.P. 2004, ApJ, 605, 819

\bibitem[Lloyd-Ronning et al. 2002]{lr02} Lloyd-Ronning, N.M., Fryer,
C.L. \& Ramirez-Ruiz, E. 2002, ApJ, 574, 554

\bibitem[M\'esz\'aros \& M\'esz\'aros 1996]{MeszarosMeszaros1996}
M\'esz\'aros, A. \& M\'esz\'aros, P. 1996, ApJ, 466, 29

\bibitem[M\'esz\'aros et al. 2006]{me06b} M\'esz\'aros, A. et al. 2006,
A\&A, 455, 785

\bibitem[M\'esz\'aros et al. 2009]{me09} M\'esz\'aros, A. et al. 2009,
Gamma-ray bursts: Sixth Huntsville Symp., AIP Conf. Proc., 1133, 483

\bibitem[M\'esz\'aros 2006]{me06} M\'esz\'aros, P. 2006,
Rep.Progr.Phys., 69, 2259

\bibitem[M\'esz\'aros \& M\'esz\'aros 1995]{MeszarosMeszaros1995}
M\'esz\'aros, P. \& M\'esz\'aros, A. 1995, ApJ, 449, 9

\bibitem[Norris 2002]{nor02} Norris, J.P. 2002, ApJ, 579, 386

\bibitem[Norris 2004]{nor04} Norris, J.P. 2004 Baltic Astronomy, 13, 221

\bibitem[Norris et al. 2000]{nor00} Norris, J.P., Marani, G.F. \&
Bonnell, J.T. 2000, ApJ, 534, 248

\bibitem[Norris et al. 2001]{nor01} Norris, J.P., Scargle, J.D. \&
Bonnell, J.T. 2001, Gamma-Ray Bursts in the Afterglow Era, eds. E.Costa,
F.Frontera, \& J.Hjorth, Springer-Verlag, 40

\bibitem[Paczy\'nski 1992]{pa92} Paczy\'nski, B. 1992, Nature, 355,
521

\bibitem[Petrosian et al. 2009]{pet09} Petrosian, V.,
Bouvier, A. \& Ryde, F. 2009, arXiv:0909.5051

\bibitem[Piran 2004]{pi04} Piran, T. 2004, Rev.Mod.Phys., 76, 1143

\bibitem[Ramirez-Ruiz \& Fenimore 2000]{rafe00} Ramirez-Ruiz, E.
\& Fenimore, E.E. 2000, ApJ, 539, 712

\bibitem[Rau et al. 2009a]{ra09a} Rau, A. et al. 2009a, GCN,
9057

\bibitem[Rau et al. 2009b]{ra09b} Rau, A. et al. 2009b, GCN,
9983

\bibitem[Reichart \& M\'esz\'aros 1997]{reme97} Reichart, D.E. \&
M\'esz\'aros, P. 1997, ApJ, 483, 597

\bibitem[Reichart D.E. et al. 2001]{rei01} Reichart, D.E. et al. 2001, ApJ, 552, 57

\bibitem[Ryde et al. 2005]{ryde05} Ryde, F. et al. 2005, A\&A, 432, 105

\bibitem[Schaefer 2003]{scha03} Schaefer B.E. 2003, ApJ, 583, L67

\bibitem[Schaefer et al. 2001]{scha01} Schaefer, B.E., Deng, M., \& Band,
D.L. 2001, ApJ, 563, L123

\bibitem[Tonry et al. 2003]{to03} Tonry, J.L. et al. 2003, ApJ, 594, 1

\bibitem[van der Horst et al. 2009]{horst09} van der Horst, A.J. et al.
2009, GCN, 9035

\bibitem[Varga et al. 2005]{var05} Varga, B. et al. 2005, Nuovo Cim. C.,
28, 861

\bibitem[Vavrek et al. 2008]{va08} Vavrek, R. et al. 2008, MNRAS,
391, 1741

\bibitem[Veres et al. 2005]{ver05} Veres, P. et al. 2005, Nuovo Cim. C.,
28, 355

\bibitem[Weinberg 1972]{we72} Weinberg, S. 1972, Gravitation and Cosmology, J.
Wiley and Sons., New York

\bibitem[Wolf \& Podsiadlowski 2007]{wp07} Wolf C. \& Podsiadlowski P. 2007, MNRAS, 375, 1049

\bibitem[Zhang et al. 2009]{zha09} Zhang, B.. et al. 2009,
ApJ, 703, 1696

\end{thebibliography}
\end{document}